\documentclass[prl,reprint,superscriptaddress]{revtex4-1}

\usepackage{graphicx}  %
\usepackage{bm}        %
\usepackage{amssymb}   %

\usepackage{color}

\begin{document}

\title{Impact of the electronic band structure in high-harmonic generation spectra of solids
}

 \author{Nicolas Tancogne-Dejean}
  \email{nicolas.tancogne-dejean@mpsd.mpg.de}
  \affiliation{Max Planck Institute for the Structure and Dynamics of Matter,               Luruper Chaussee 149, 22761 Hamburg, Germany}
 \affiliation{European Theoretical Spectroscopy Facility (ETSF)}

 \author{Oliver D. M\"ucke}
 \affiliation{Center for Free-Electron Laser Science CFEL, \\Deutsches Elektronen-Synchrotron DESY, Notkestra\ss e 85, 22607 Hamburg, Germany}
\affiliation{The Hamburg Center for Ultrafast Imaging, Luruper Chaussee 149, 22761 Hamburg, Germany}

 \author{Franz X. K\"artner}
  \affiliation{Center for Free-Electron Laser Science CFEL, \\Deutsches Elektronen-Synchrotron DESY, Notkestra\ss e 85, 22607 Hamburg, Germany}
\affiliation{The Hamburg Center for Ultrafast Imaging, Luruper Chaussee 149, 22761 Hamburg, Germany}
 \affiliation{Physics Department, University of Hamburg, Luruper Chaussee 149, 22761 Hamburg, Germany}
 \affiliation{Research Laboratory of Electronics, Massachusetts Institute of Technology, 77 Massachusetts Avenue, Cambridge, MA 02139, USA}

 \author{Angel Rubio}
  \email{angel.rubio@mpsd.mpg.de}
\affiliation{Max Planck Institute for the Structure and Dynamics of Matter,               Luruper Chaussee 149, 22761 Hamburg, Germany}
 \affiliation{European Theoretical Spectroscopy Facility (ETSF)}
 \affiliation{Center for Free-Electron Laser Science CFEL, \\Deutsches Elektronen-Synchrotron DESY, Notkestra\ss e 85, 22607 Hamburg, Germany}
\affiliation{Physics Department, University of Hamburg, Luruper Chaussee 149, 22761 Hamburg, Germany}

\begin{abstract}

An accurate analytic model describing high-harmonic generation (HHG) in solids is derived. Extensive first-principles simulations within a time-dependent density-functional framework corroborate the conclusions of the model. Our results reveal that: (i) the emitted HHG spectra are highly anisotropic and laser-polarization dependent even for cubic crystals, (ii) the harmonic emission is enhanced by the inhomogeneity of the electron-nuclei potential, the yield is increased for heavier atoms, and (iii) the cutoff photon energy is driver-wavelength independent. Moreover, we show that it is possible to predict the laser polarization for optimal HHG in bulk crystals solely from the knowledge of their electronic band structure. Our results pave the way to better control and optimize HHG in solids by engineering their band structure.

\end{abstract}

\maketitle

Atoms and molecules interacting with strong laser pulses emit high-order harmonics of the fundamental driving laser field.
The high-harmonic generation (HHG) in gases is routinely used nowadays to produce isolated attosecond pulses~\cite{drescher2001x, Zhao_OL12, Takahashi_NC13,  hammond2016attosecond} and coherent radiation ranging from the visible to soft X-rays \cite{Popmintchev_Sci12}.
Due to a higher electronic density, solids are one promising route towards compact, brighter HHG sources. The recent observation of non-perturbative HHG in solids without damage \cite{Ghimire2011,Schubert2014,Hohenleutner2016,Luu2015,ndabashimiye2016solid}, extending even beyond the atomic limit~\cite{ndabashimiye2016solid}, has opened the door to the observation and control of attosecond electron dynamics in solids \cite{Muecke_PRB11,Luu2015, Hohenleutner2016}, all-optical band-structure reconstruction \cite{Vampa2015PRL}, and solid-state sources  of isolated extreme-ultraviolet (XUV) pulses  \cite{Muecke_PRB11,Luu2015}.
However, in contrast to HHG from gases, the microscopic mechanism underlying HHG from solids is still controversially debated in the attoscience community, in some cases casting doubts on the validity of the proposed microscopic model and resulting in confusion about the correct interpretation of experimental data. 
Various competing simplified models have been proposed but they often are based on strong approximations and {\it a priori} assumptions, often stating that there is a strong similarity with the processes underlying atomic-gas HHG emission.
However, it is clear that many-body effects due to the crystalline structure of solids and the fermionic nature of interaction electrons play a decisive role that fundamentally distinguishes the solid from the gas case.
It is the scope of the present work, to unravel within an {\it ab initio} approach, what is the impact of the underlying electronic band structure of the solids in the observed HHG emission.

The process of HHG from gases is by now well understood in terms of the three-step model~\cite{PhysRevLett.70.1599,PhysRevLett.71.1994,kuchiev1987atomic} in which electrons are first promoted from the ground state of the atom (or molecule) to the continuum, then accelerated by the electric field and finally recombine with the parent ion.
With this simple, intuitive model most of the observed effects are well described, in particular the dependence of the harmonic cutoff energy on driver wavelength and intensity. In the case of solids, electrons are promoted to discrete conduction bands, where they do not evolve freely. This leads, for instance, to a different linear field dependence for the cutoff energy~\cite{Ghimire2011}, different  time-frequency characteristics of the harmonic emission between atoms and solids~\cite{Muecke_PRB11,Huttner2016, Hohenleutner2016}, and a different ellipticity dependence~\cite{Ghimire2011,You16}.
As we will show below, the wavelength dependence of the cutoff energy is also different.

Historically, HHG in solids was first discussed in terms of Bloch oscillations (i.e., pure intraband dynamics)~\cite{Ignatov_pssb, Feise_APL, Wegener_2005}, and more recently mainly analyzed using simplified models based on numerical solutions of the semiconductor Bloch equations~\cite{Golde_PRB,Golde_PSSC,Golde_PSSB} treating the complex, coupled interband and intraband dynamics.
Even if these methods have been successfully applied to some materials, such as GaSe~\cite{Schubert2014,Hohenleutner2016} or SiO$_2$~\cite{Luu2015}, basic questions remain controversial and/or unresolved, e.g,  which bands are
involved in the HHG dynamics~\cite{Hawkins2015,Wu2015}, why can HHG in the covalently bonded crystals ZnO and GaSe be well described by intraband Bloch oscillations~\cite{Ghimire2011,Muecke_PRB11, Schubert2014}, whereas the van-der-Waals bonded, rare-gas solids Ar and Kr are better described by four-level Bloch equations~\cite{ndabashimiye2016solid}.

The first experimental observations of HHG from solids were explained in terms of Bloch oscillations~\cite{Ghimire2011,Schubert2014}.
It was shown in~\cite{Ghimire2011} that the non-parabolicity of the conduction band dispersion was sufficient to produce high-order harmonics.
Moreover, a competing model attributing the HHG mechanism to interband transitions (resembling the three-step model of gas HHG \cite{PhysRevLett.70.1599,PhysRevLett.71.1994}) was introduced~\cite{Vampa2014,vampa2015linking}. For not too strong excitation of the semiconductor ZnO with mid-infrared pulses (from 2 to 6$\mu$m), such that the electrons explore only the near-parabolic region of the Brillouin zone (BZ), 
it was found that the magnitude of the interband contribution is larger than that of the intraband contribution~\cite{Vampa2014}.  
Nevertheless, most theoretical works have used either a two-band or a five-band model, intrinsically hampering the predictive power of the model and the full microscopic understanding of the HHG process.

In this letter, using an \textit{ab initio} approach based on time-dependent density-functional theory (TDDFT)~\cite{PhysRevLett.52.997, PhysRevLett.80.1280}, we study the microscopic origin of HHG in solids.
Effects stemming from the full electronic structure (valence and conduction bands) and the real crystal structure are properly accounted for.
We show that the non-perturbative emission of harmonics in solids arises from the interplay between intraband and interband contributions, and most importantly, that it can be enhanced when the interband contribution is suppressed due to band-structure effects. We identify that the {\it joint density of states (JDOS) along the laser polarization} is the key parameter governing the weight of the interband contribution. Knowledge of this JDOS permits the prediction of the optimal laser polarization for harmonic generation, and thus might pave the way to control and enhance HHG from solids by tailoring their band structure.
In addition, we address the still controversial question of the wavelength dependence of the cutoff energy in bulk crystals. The cutoff of HHG in solids was proposed to be wavelength-independent~\cite{Wegener_2005}. However, some recent theoretical studies found it to depend linearly on the wavelength~\cite{Wu2015,Guan2016,Vampa2015}, in contrast to the conclusions of other works~\cite{Ghimire2011,Muecke_PRB11, Ghimire2012,Higuchi2014,Luu2015}.
Here, we confirm with our first-principles simulations that the cutoff energy is indeed independent of the laser wavelength in solids (as proposed in Ref.~\cite{Wegener_2005}), in contrast to what is found in the atomic and molecular case, where the ponderomotive energy $U_{\rm p}\propto \lambda^2 I$ and consequently the cutoff energy depend quadratically on the laser wavelength.

We start by presenting some exact analytical results.
We consider a general interacting many-electron Hamiltonian $\hat{H}$ of the form
\begin{equation}
\hat{H}(t) = \hat{T} + \hat{V}(t) + \hat{W},
\label{eq:hamiltonian}
\end{equation}
where $\hat{T}$ is the kinetic energy, $\hat{V}(t)$ is the time-dependent external laser potential, and $\hat{W}$ is the electron-electron Coulomb interaction (the ionic motion is not considered here for the sake of simplicity).
The exact equation of motion for the total microscopic current, $\mathbf{j}(\mathbf{r},t)$, can be rewritten as~\cite{PhysRevLett.82.3863,stefanucci2013nonequilibrium}
\begin{equation}
\frac{\partial}{\partial t}\mathbf{j}(\mathbf{r},t) = -n(\mathbf{r},t) \nabla v(\mathbf{r},t) + \Pi^{\mathrm{kin}}(\mathbf{r},t) + \Pi^{\mathrm{int}}(\mathbf{r},t),
\label{eq:third_newtons_law}
\end{equation}
where $\Pi^{\mathrm{kin}}(\mathbf{r},t)$ and $\Pi^{\mathrm{int}}(\mathbf{r},t)$ are the kinetic and the interaction contributions to the momentum-stress tensor~\cite{PhysRevLett.82.3863, Momentum_stress_tensors,stefanucci2013nonequilibrium}. This equation just represents the local momentum conservation law, and shows that only external forces contribute to the total momentum, in accordance to Newton's third law.
As these two contributions to the momentum stress-tensor are internal forces~\cite{stefanucci2013nonequilibrium}, Eq.~(\ref{eq:third_newtons_law}) reduces to
\begin{equation}
  \frac{\partial}{\partial t} \int_{\Omega} d^3\mathbf{r}\, \mathbf{j}(\mathbf{r},t) = -\int_{\Omega} d^3\mathbf{r}\,n(\mathbf{r},t) \nabla v(\mathbf{r},t),
  \label{eq:total_current}
\end{equation}
where $\Omega$ denotes the volume of the physical system. Eq.~(\ref{eq:total_current}) provides an \emph{exact} relation, valid for atoms, molecules as well as solids, that allows us to obtain a new formula for the high-harmonic spectra.
Using the current expression for the HHG spectra, namely $\mathrm{HHG}(\omega) = \left|\mathrm{FT}\left\{\int_{\Omega} d^3\mathbf{r}\, \frac{\partial}{\partial t}\mathbf{j}(\mathbf{r},t)\right\}\right|^2$,
and plugging now Eq.~(\ref{eq:total_current}), we obtain a general expression for the HHG spectra
\begin{eqnarray}
\mathrm{HHG}(\omega) &\propto& \Bigg|\mathrm{FT}\Bigg\{\int_{\Omega} d^3\mathbf{r}\,\Big(n(\mathbf{r},t) \nabla v_0(\mathbf{r}) \nonumber\\
 &&+ n(\mathbf{r},t)\mathbf{E}(\mathbf{r},t)+ \frac{\mathbf{j}(\mathbf{r},t)\times\mathbf{B}(\mathbf{r},t)}{c}\Big)\Bigg\}\Bigg|^2,
\label{eq:new_formula_HHG1}
\end{eqnarray}
where the last two terms correspond to the Lorentz force exerted by the external laser on the electronic system~\cite{stefanucci2013nonequilibrium}.
If we now make the dipole approximation, Eq.~(\ref{eq:new_formula_HHG1}) further simplifies and we finally get
\begin{equation}
\mathrm{HHG}(\omega) \propto \left|\mathrm{FT}\Bigg\{\int_{\Omega} d^3\mathbf{r}\, n(\mathbf{r},t) \nabla v_0(\mathbf{r}) \Bigg\} + N_e\mathbf{E}(\omega) \right|^2,
\label{eq:new_formula_HHG}
\end{equation}
which provides the first important physical result of the present work, shedding fundamental insights on the intrinsic bulk contribution to the HHG spectra.
In here the external potential $v(\mathbf{r},t)$ accounts for both the electron-nuclei potential ($v_0(\mathbf{r})$) and the externally applied time-dependent laser field.
$n(\mathbf{r},t)$ is the time-dependent electronic density of the system driven by the external strong laser pulse $\mathbf{E}(t)$ thereby generating the higher harmonics, and $N_e$ is the number of electrons contained in the volume $\Omega$. Note that the HHG spectra depend only on the electronic density.
The second term does not result in a non-perturbative nonlinearity and thus can not create a plateau-like HHG spectrum. 
The more interesting and relevant term for HHG is the first one in Eq.~(\ref{eq:new_formula_HHG}). It shows that higher harmonics are generated by two competing terms, the spatial variation of the total electronic density ($n(\mathbf{r},t)$) and the gradient of the electron-nuclei potential ($\nabla v_0(\mathbf{r})$), the latter being time independent, as we neglected ionic motion\footnote{The contribution of vibrations to the HHG spectra is left for a forthcoming work.}. In gases, the gradient of the electron-nuclei potential is important, but the electronic density is low.
In the case of solids, the electronic density is higher, but the potential is rather homogeneous, resulting in a smaller gradient of the potential than in the atomic case. In fact, in the limit of a homogeneous electron gas, the gradient becomes zero, and no harmonics are generated, irrespective of the value of the electronic density. In this case the bands are parabolic, thus we recover the known result that parabolic bands do not yield non-perturbative harmonics~\cite{Ghimire2011}.

Since the gradient electron-nuclei potential is frequency-independent, it contributes equally to all harmonics equally and therefore could be used to enhance the entire HHG spectra. As consequence, we expect a higher harmonic yield when we have strong spatial fluctuations of the electron-nuclei potential, as can be realized at surfaces or interfaces. This also means that layered materials, such as transition-metal dichalcogenides (TMD)~\cite{mak2016photonics}, should be good candidates for HHG. Finally, we note that a similar expression, valid only for atoms, was obtained in Ref.~\cite{PhysRevLett.96.223902}. Their equation (9) was used to explain the dependence of HHG yield on atomic number $Z$ for noble gases only. Here, we suggest that the yield of HHG in solids will also increase with the atomic number as in atoms. This corroborates the idea that layered TMD are good candidates for improving the yield of HHG.
These results might therefore guide the search of better materials for HHG from solids, as not only bulk crystal properties but also nanostructure engineering aspects are important for optimum HHG.

Next we discuss the numerical results of our first-principles TDDFT calculations. Being interested in the microscopic origin of HHG in solids, we have neglected macroscopic propagation effects in our quantum-mechanical simulations, thus making a sudden approximation, and we consider only the intrinsic bulk contribution. 
As most previous works have described HHG from solids in terms of the dynamics of non-interacting electrons, we explore here how the Coulomb interaction and electron-electron correlations affect the HHG in solids. We consider a laser pulse of 25-fs duration, with a sin-square envelope of the vector potential. The peak intensity inside matter is taken to be $I_0=10^{11}$W\,cm$^{-2}$, (see supplementary material for higher intensity) and the carrier wavelength $\lambda$ is 3000\,nm, corresponding to a carrier photon energy of 0.43\,eV. For such few-cycle driver pulses, the HHG spectra from solids have been shown to be quite insensitive to the carrier-envelope phase (CEP)~\cite{Muecke_PRB11,Kemper2013,Luu2015}, which is therefore taken to be zero here.
The evolution of the wave-functions and the evaluation of the time-dependent current is computed by propagating the Kohn-Sham equations within TDDFT, as provided by the Octopus package~\cite{C5CP00351B}, in the local-density approximation (LDA).

Simulations are performed for the prototype system bulk silicon \footnote{All calculations were performed using the primitive cell of bulk silicon, using a real-space spacing of 0.484 atomic units, and an optimized 28$\times$28$\times$28 grid shifted four times to sample the BZ. We employ norm-conserving pseudo-potentials, and use the experimental lattice constant. The LDA band gap of silicon of 2.58\,eV corresponds to six times the carrier photon energy 0.43\,eV for 3000-nm driver pulses. JDOS are computed using the ABINIT software~\cite{ABINIT} and the DP code (V. Olevano, L. Reining, and F. Sottile; http://dp-code.org). HHG spectra are obtained from the power spectra of the time derivative of the current.} (and AlAs, see supplementary material), which exhibits a richer and more complex band structure close to the Fermi energy than previously studied materials, such as ZnO~\cite{Ghimire2011,Muecke_PRB11}, GaSe~\cite{Schubert2014,Hohenleutner2016} and SiO$_2$~\cite{Luu2015}. 
Moreover, it is highly relevant for semiconductor technology. It is therefore our material of choice for investigating the origin of HHG in solids.

\begin{figure}[t]
  \begin{center}
    \includegraphics[width=\columnwidth]{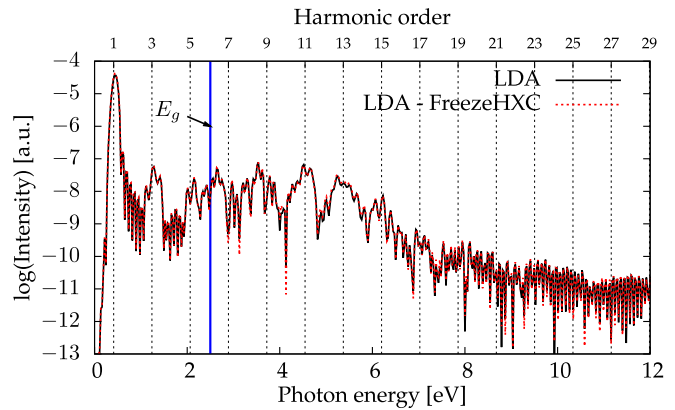}
  \end{center}
  \caption{\label{img:correlations} HHG spectra from bulk silicon, for polarization along $\overline{\Gamma X}$, computed within the LDA (LDA; black line) and within the LDA, but freezing the Coulomb and exchange-correlation terms to their ground-state value (LDA-FreezeHXC; red line). }
\end{figure}

From Fig.~\ref{img:correlations} we find that the HHG spectrum of bulk silicon does not change if we consider either the full evolution of the Hartree and the exchange-correlation parts of the Kohn-Sham Hamiltonian or the time evolution in a static ground-state potential. This means that, in silicon, electrons evolve mainly as independent particles in the ground-state potential for our excitation conditions. In the language of atomic HHG, this is similar to the widely used single-active electron (SAE) approximation.
This result has two important implications: First, it justifies the independent-particle approximation assumed in most previously published HHG models. Second, it implies that ground-state information of the crystal, such as the band structure, might be retrieved from the HHG spectra. \footnote{This result was obtained for silicon and it might not be true in general, and other materials could exhibit strong correlation effects, e.g., as observed for atomic HHG from Xe \cite{Shiner2011}.} Moreover, the band-structure information could be altered by light propagation effects in the solid, possibly including impurity or lattice scattering. Only a careful analysis of these extrinsic effects, well beyond the scope of this paper, could show how harmonics are affected while accounting for propagation effects.

\begin{figure}[t]
  \begin{center}
    \includegraphics[width=0.9\columnwidth]{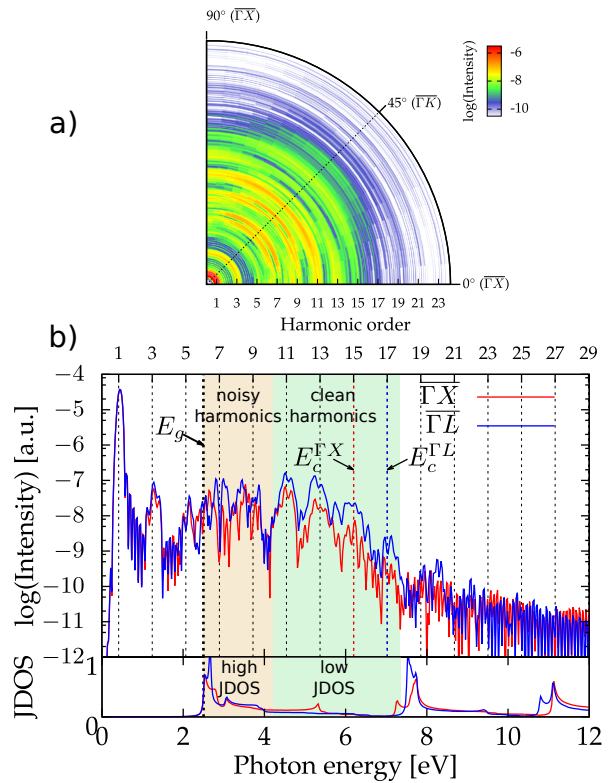}
  \end{center}
  \caption{\label{img:JDOSvsHHG} a) Calculated TDDFT anisotropy map of the HHG spectra obtained by rotating the laser polarization around the [001] crystallographic direction, from $0^{\circ}$ (along $\overline{\Gamma X}$) to $45^{\circ}$ ($\overline{\Gamma K}$) to $90^{\circ}$ ($\overline{\Gamma X}$).  b) HHG spectra for the $\overline{\Gamma X}$ polarization direction (red line) and the $\overline{\Gamma L}$ direction (blue line). The bottom panel shows the corresponding joint density of states (JDOS). The red and blue dashed lines indicate the position of the cutoff energy ($E_c$) for  $\overline{\Gamma X}$ and $\overline{\Gamma L}$ directions, respectively. The shaded areas are guides to the eye.}
\end{figure}

We now investigate the effect of the laser polarization on the HHG emission.
For sake of simplicity, let us consider a general cubic material. In this case, the laser electric field is driving the electrons along the direction of the laser polarization. Orienting the laser polarization along specific directions, corresponding to high-symmetry lines of the 3D BZ of the crystal, thus results in different HHG spectra. Therefore, even cubic materials such as silicon, will exhibit a strong anisotropic emission of high-order harmonics. Moreover, the symmetries of the crystal, which are also the symmetries of the BZ, are reflected in the anisotropy of the HHG emission.
Our simulation results, displayed in Fig.~\ref{img:JDOSvsHHG}a, clearly predicts an anisotropic emission of harmonics while rotating the polarization around the [001] crystallographic axis. The harmonic emission is maximum for a laser polarization along the $\overline{\Gamma K}$ direction, and it is significantly reduced for the $\overline{\Gamma X}$ direction.

Considering the mechanism underlying HHG in solids, we first note that harmonics emitted energetically below the band-gap energy can not originate from the recombination of an electron with a hole present in the valence bands, as this would lead to the emission of a photon with energy above the band-gap energy. This indicates that below-band-gap harmonics cannot originate from the interband contribution. In other words, below the band gap, the interband emission channel is naturally suppressed. This is the case in experiments performed on bulk GaSe~\cite{Schubert2014,Hohenleutner2016}, for which the numerical calculations reproduce quite well the clean shape of the harmonic peaks observed in the experiments ~\cite{Schubert2014}, and the temporal profile of harmonic emission~\cite{Hohenleutner2016}.

Above the band gap, in contrast, it becomes energetically possible that emitted harmonics originate from an interband electron-hole recombination.
In this situation, both interband and intraband dynamics contribute to harmonics emitted above the band gap.
Interestingly, clear above-band-gap odd-harmonic peaks have been observed experimentally in ZnO ~\cite{Ghimire2011}, whereas the above-band-gap plateau has been found theoretically to be strongly modulated~\cite{Huttner2016,Vampa2014,Kemper2013,Wu2015,Guan2016,Vampa2015,tamaya2016diabatic}.
The absence of clean harmonics in the theoretical works has previously been attributed to an infinitely long dephasing time, when considering a two-band model~\cite{Vampa2014}, to a metallization regime~\cite{tamaya2016diabatic}, or to symmetry breaking for a three-band model~\cite{Huttner2016}. A study performed using non-equilibrium Green's functions~\cite{Kemper2013} has shown that a noisy plateau could originate also from elastic or inelastic scattering processes. Finally, we point out that such strongly modulated plateau cannot originate here from inter-cycle or intra-cycle interferences, as observed in above-threshold ionization (ATI) from gases \cite{PhysRevA.81.021403, PhysRevLett.108.193004}, because  such interferences would affect the entire HHG spectra, and not only the above-band-gap region.

The emission of harmonics by interband transitions in solids is naturally dictated by the discretization of the bands in solids. This represents one of the biggest differences between atomic/molecular HHG and the HHG in solids. In order to emit a photon at a given energy by interband transitions, the corresponding direct transition must be possible between two states.
The density of possible transitions at a given energy, namely the JDOS, is thus intrinsically related to the interband mechanism. More precisely, it is the JDOS corresponding to the region of the BZ explored by the electrons which dictates the emission of harmonics by interband transitions.

Similarly to previous theoretical studies~\cite{Huttner2016,Vampa2014,Kemper2013,Wu2015,Guan2016,Vampa2015,tamaya2016diabatic}, we do not obtain clean odd harmonics above the band gap.
Nevertheless, we see in Fig.~\ref{img:JDOSvsHHG}b) that the noisy region (orange shaded area) is suppressed, thus recovering clean odd harmonics (green shaded area), when the JDOS (computed for the region explored by the electrons, assuming the acceleration theorem) is very low, corresponding to the situation when the electron-hole recombination channel is drastically reduced.
Interestingly, we observe that selecting the laser polarization along the $\overline{\Gamma L}$ high-symmetry line leads to generation of harmonics up to the 17th harmonic, whereas only the first 15 harmonics are generated when the laser polarization is set along the $\overline{\Gamma X}$ high-symmetry line.
Moreover, the 13th and 15th harmonics are more intense for the $\overline{\Gamma L}$ case compared to the $\overline{\Gamma X}$ spectrum (see Fig.\ref{img:JDOSvsHHG}b)).
This suggests that more intense and energetic harmonics are obtained when suppressing interband transitions. Therefore, with knowledge of the ground-state JDOS, a direct prediction of the optimal laser polarization for HHG in solids is possible. This also paves the way to control and improvement of the yield of HHG in solids via band-structure engineering, for instance by opening gaps between conduction bands.

\begin{figure}[t]
  \begin{center}
    \includegraphics[width=0.9\columnwidth]{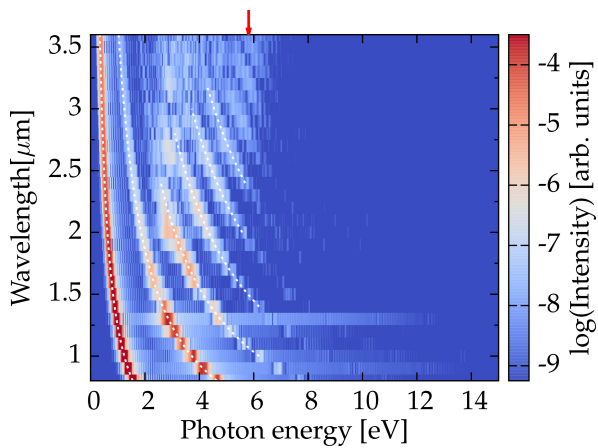}
  \end{center}
  \caption{\label{img:HHGvsWavelength} HHG spectra versus center wavelength of the driver pulses, at fixed peak intensity and laser pulse duration, for polarization along $\overline{\Gamma X}$. White dashed curves represent the harmonics and the red arrow indicates the wavelength-independent harmonic cutoff.}
\end{figure}

We finally address a fundamentally and technologically relevant aspect of the emission of HHG, which is the wavelength dependence of the cutoff photon energy in harmonic spectra. Much research effort has been devoted to identify key parameters governing the HHG cutoff energy. Surprisingly, the wavelength scaling of the cutoff energy is still not clearly established theoretically, as some studies found it to be wavelength-independent~\cite{Ghimire2011, Muecke_PRB11,Ghimire2012,Higuchi2014,Luu2015}, whereas others claimed it depends linearly on the wavelength~\cite{Wu2015,Guan2016,Vampa2015}. 
Our \textit{ab initio} quantum-mechanical simulations displayed in Fig.~\ref{img:HHGvsWavelength} confirm that the HHG cutoff energy is independent of the driver laser wavelength.

In gases, the wavelength dependence comes from the ponderomotive energy $U_{\rm p}\propto \lambda^2 I$, which originates from the {\it free} evolution of the ionized electron in the continuum accelerated by the laser field.
In the case of solids, it is clear that electrons do not evolve as free particles. Thus, for solids, a wavelength dependence cannot arise from the ponderomotive energy.
We note also that increasing the wavelength, clear perturbative harmonics disappear in Fig.~\ref{img:HHGvsWavelength} in a white-noise-type plateau, characteristic of a non-perturbative regime.
The wavelength independence of the cutoff energy offers great technological perspectives, as it permits a greater flexibility in the choice of the driver laser pulses.

In conclusion, we analyzed the microscopic origin of high-harmonic generation in solids. We show analytically that high-harmonic generation in solids is enhanced by the inhomogeneity of the electron-nuclei potential, and that the yield is increased when we have heavier atoms in the solid. Our \textit{ab initio} simulations demonstrate that HHG in bulk crystals is anisotropic, even in cubic materials. 
Our simulations revealed that it is possible to suppress interband transitions in favor of HHG arising from intraband dynamics in solids, and most importantly to predict the optimal laser polarization, based on the sole knowledge of the crystal’s band structure and its JDOS. Finally, we confirmed without making any model assumptions that the cutoff energy of the HHG in solids is wavelength-independent, offering many intriguing technological perspectives.
Further investigations should address extrinsic effects such as the electron-phonon coupling, propagation and surface effects.
We expect this work will help in the search of better materials for solid-state high-harmonic sources and tailored HHG in solids.

\begin{acknowledgments}
We acknowledge financial support from the European Research Council (ERC-2015-AdG-694097), COST Action MP1306 (EUSpec).
N.T.-D. and A.R. would like to thank K.-M. Lee, S.A. Sato and T.J.-Y. Derrien for helpful discussions.
F.X.K. and O.D.M. acknowledge support by the excellence cluster 'The Hamburg Centre of Ultrafast Imaging-Structure, Dynamics and Control of Matter at the Atomic Scale' and the priority program QUTIF (SPP1840 SOLSTICE) of the Deutsche Forschungsgemeinschaft.
\end{acknowledgments}

\bibliography{bibliography}

\begin{thebibliography}{48}
\bibitem [{Mom()}]{Momentum_stress_tensors}%
  \BibitemOpen
  \href@noop {} {}\bibinfo {note} {In second quantization, the $k$-component of
  $\Pi^{\mathrm{kin}}(\mathbf{r},t)$ and $\Pi^{\mathrm{int}}(\mathbf{r},t)$ are
  given respectively by $\Pi^{\mathrm{kin}}_k(\mathbf{r},t) = \langle
  \Psi(t)|\frac{1}{2}\sum_i \partial_i\{
  \partial_i\hat{\psi}^\dag(\mathbf{r})\partial_k\hat{\psi}(\mathbf{r}) +
  \partial_k\hat{\psi}^\dag(\mathbf{r})\partial_i\hat{\psi}(\mathbf{r}) -
  \frac{1}{2}
  \partial_i\partial_k[\hat{\psi}^\dag(\mathbf{r})\hat{\psi}(\mathbf{r})]
  \}|\Psi(t)\rangle$ and $\Pi^{\mathrm{int}}_k(\mathbf{r},t) = \langle
  \Psi(t)|\int d^3\mathbf{r'}
  \hat{\psi}^\dag(\mathbf{r})\hat{\psi}^\dag(\mathbf{r'})\partial_k\,w(|\mathbf{r}-\mathbf{r'}|)\hat{\psi}(\mathbf{r'})\hat{\psi}(\mathbf{r})|\Psi(t)\rangle$,
  where $|\Psi(t)\rangle$ is the state evolving from the initial state
  $|\Psi_0\rangle$ under the influence of $\hat{H}(t)$.}\BibitemShut {Stop}%
\end{thebibliography}

\end{document}